\documentclass[10pt,pra,twocolumn,superscriptaddress]{revtex4}

\usepackage{graphicx}
\usepackage{amssymb}
\usepackage{times}
\usepackage{amsmath}
\usepackage{amsfonts}
\usepackage{txfonts}
\usepackage{color}

\begin{document}

\title{Cavity opto-electromechanical system combining strong electrical actuation with ultrasensitive transduction}

\author{Terry G. McRae} \affiliation{Department of Physics, University of Queensland, St Lucia,
Queensland 4072, Australia} \affiliation{MacDiarmid Institute, Physics Department, University of Otago, Dunedin, New Zealand}

\author{Kwan H. Lee} \affiliation{Department of Physics, University of Queensland, St Lucia,
Queensland 4072, Australia}

\author{Glen I. Harris} \affiliation{Department of Physics, University of Queensland, St Lucia,
Queensland 4072, Australia}

\author{Joachim Knittel} \affiliation{Department of Physics, University of Queensland, St Lucia,
Queensland 4072, Australia}

\author{Warwick P. Bowen} \affiliation{Department of Physics, University of Queensland, St Lucia,
Queensland 4072, Australia}

\begin{abstract}
 A cavity opto-electromechanical system is reported which combines the ultrasensitive transduction of cavity optomechanical systems with the electrical actuation of nanoelectromechanical systems. Ultrasensitive mechanical transduction is achieved via opto-mechanical coupling. Electrical gradient forces as large as 0.40~$\mu$N are realized, facilitating strong actuation with ultralow dissipation. A scanning probe microscope is implemented, capable of characterizing the mechanical modes. The integration of electrical actuation into optomechanical devices is an enabling step towards the regime of quantum nonlinear dynamics, and provides new capabilities for quantum control of mechanical motion.
\end{abstract}


\date{\today} \maketitle

\section{Introduction}
Mechanical oscillators offer the potential to test quantum mechanics at the macroscopic scale\cite{Schwab05,Rugar2,Milburn,Katz,Bouwmeester03,OConnell2010}, as well as significant technological advances including applications in metrology\cite{Rugar} and quantum information systems\cite{Mancini,Rabl}. Two key requirements to enter this quantum regime are the capacities to efficiently quantum control the oscillator, and to achieve zero-point motion limited transduction sensitivity. Strong electrical actuation of mechanical oscillators, an enabling step for quantum control, has been achieved in nanoelectromechanical systems (NEMS)~\cite{Lehnert,Schwab,Poggio,Blair95,Hertzberg2009,Unterreithmeier09}. Furthermore, transduction sensitivities near the quantum limit have been reached with cavity optomechanical systems (COMS)\cite{Cleland09,Kippenberg09}. We have recently combined these two techniques into the first cavity opto-electromechanical system (COEMS)\cite{Lee2010}; achieving gradient forces an order of magnitude stronger than any cryogenic radiation pressure based microcavity actuation technique to date, with intracavity optical enhancement enabling transduction at levels close to the mechanical zero-point motion\cite{Kippenberg09}. In this paper we present further analysis of this system including new theoretical and experimental results confirming gradient forces as the mechanism responsible for mechanical actuation, an explicit derivation of the peak to peak gradient force and new results imaging mechanical modes with COEMS.

The capacity to strongly electrically actuate COMS reported here has several important ramifications. Electrical feedback of the optomechanical transduction signal allows immediate control of the properties of the mechanical oscillator;  facilitating, for example, active cooling or heating\cite{Lee2010,Cohadon,Pinard2000}, electro-optic spring effects\cite{Vahala}, and  phonon lasing\cite{Vahala2}. Strong electrical driving of the mechanical motion provides access to mechanical nonlinear dynamics in both classical and ultimately quantum regimes; with nanofabrication techniques providing the means to engineer the nonlinear properties. This nonlinear mechanical regime is usually achieved with NEMS. For example the coherent response of mechanical oscillations allow signal enhancement and noise reduction\cite{Cross}, ultrasensitive mass detection\cite{Spletzer,Sato}, sub-Heisenberg limit metrology\cite{Milburn}, and mechanical quantum state engineering\cite{Katz}. The superior transduction sensitivity afforded by COMS presents an enabling step towards observation of quantum nonlinear dynamics in these systems. Furthermore, the integration of electrical actuation into COMS advances the possibility of incorporating ultracold superconducting circuits, and unifying cavity opto-mechanics with the rapidly developing field of circuit quantum electrodynamics\cite{Girvin08}.

COMS consist of a high quality optical cavity strongly coupled to a mechanical oscillator.
Motion of the mechanical oscillator alters the cavity's optical path length, and as a result this motion is transferred to the out-coupled optical field. Information extracted from measurements of this field enables both direct {\it characterization} of the mechanical oscillator, and, combined with mechanical actuation, {\it control} of its behaviour\cite{Cohadon,Pinard2000}. To date, the actuation has been achieved using radiation pressure from an amplitude modulated optical field (for example see Refs.~\cite{Cohadon,Pinard2000,Kleckner}). However, radiation pressure is an inherently weak force. Furthermore, optical heating via optical absorption of intracavity power severely constrains the possibility of operation in the quantum regime\cite{Kippenberg09}. By contrast, electrical actuation of NEMS is inherently strong, much less prone to heating effects, and comparatively straightforward to implement\cite{Schwab,Poggio,Blair95,Hertzberg2009,Roukes08}.

\begin{figure}[b!]
\begin{center}
\includegraphics[width=8.5cm]{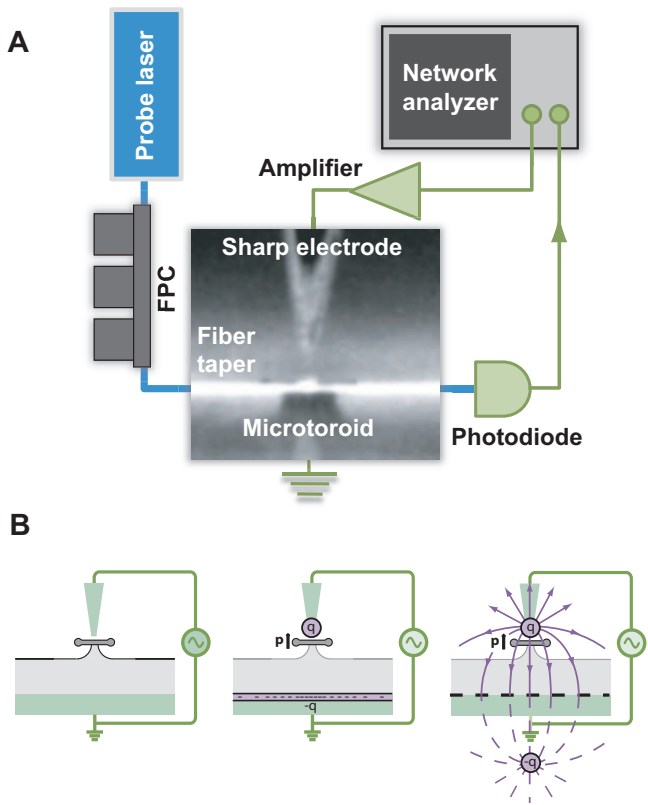}
\caption{(color online) {\bf A} Schematic of gradient force actuation experiment. FPC: fibre polarization controller. Inset:  electric field distribution between sharp and flat electrodes. {\bf B} The method of images used to describe the electric field between the sharp and flat electrodes. Left to right: mechanical and electrical structure; structure showing induced charges in sharp and flat electrodes, and microtoroid polarization; structure showing electric field lines modeled using the method of images where the flat electrode can be modeled as an equivalent but opposite sign charge to that on the electrode tip located equidistant beneath the flat electrode itself.}
\label{expt}
\end{center}
\end{figure}

NEMS are submicron structures that use actuators to convert electrical energy to mechanical energy by exciting a resonant mode structure. These structures have fundamental resonance frequencies in the radio frequency (RF) to microwave frequency domain, mechanical quality factors ,$Q_m$, in the hundreds of thousands and masses in the femtogram range\cite{Ekinci05}. External actuation mechanisms that use dieletric gradient forces reduce constraints on material choice and system geometry, and also avoid dissipative actuation losses\cite{Unterreithmeier09}. These dieletric gradient forces are the result of a non-uniform applied electric field that polarizes the dielectric so that it experiences a dipole force in the direction of increasing electric field. NEMS, however, have relatively poor transduction sensitivity on the order of $10^{-15}$~m~Hz$^{-1/2}$\cite{Knobel,Flowers-Jacobs,LaHaye,Etaki} in comparison to the COMS mentioned above which achieve transduction sensitivities better than $10^{-18}$~m~Hz$^{-1/2}$\cite{Kippenberg09}. The COEMS presented in Ref.\cite{Lee2010} integrates dieletric gradient force actuation with silica microtoroiods on a silicon chip which have both high quality mechanical resonances and ultra high quality optical resonances. This simultaneously allows gradient forces as large as 0.4$\mu$N, and  ultrasensitive transduction at the level of $1.5 \! \times \! 10^{-18}$~m~Hz$^{-1/2}$.

To demonstrate the broader relevance of this work, we implement COEMS-based scanning probe microscopy of the microtoroid mechanical mode structure. This is a natural extension of the ability to both precisely actuate and measure the displacement of the resonator. Microscopy based on the  measurement of a displacement of some spring structure in response to a force is now common-place. Atomic force microscopy (AFM)\cite{binning} is able to produce three dimensional images with atomic scale resolution of both conductors and insulators. The force on the probe may be due to a range of proximity forces including van der Walls, capillary, electrostatic, and Casimir forces; and is kept small and constant with feedback mechanisms\cite{Hutter1994,Mohideen1998}. In another example, magnetic resonance force microscopy is based on the detection of the magnetic force between a ferromagnetic tip and the electron spins in a sample and can detect single electron spins\cite{rugar2004}.
Of more direct relevance to the work reported here, the spatial profile of acoustic modes in high finesse Fabry-P\'{e}rot cavities have been imaged where the actuation force has been achieved via both optical radiation pressure\cite{Briant2003} or electrostatic actuation\cite{Arcizet2008}. In this paper we extend these techniques to measure a two dimensional spatial profile in integrated microtoroid cavities. Although currently limited in resolution due to the size of the region which our actuation force is applied to; this technique allows confirmation of the mechanical mode shapes of the microtoroid and gives a spatial profile that is in agreement with finite element modeling (COMSOL Multiphysics). With the implementation of feedback similar to that used in AFM to control the probe position sub-micron resolution should be possible.

\section{Experiment}
Figure~\ref{expt}A shows a schematic of our experimental setup. The microtoroids were fabricated on a silicon wafer with a 2 $\mu$m layer of wet grown silica. The silica was patterned using standard photolithography techniques, followed by etching in hydrofluoric acid to transfer a pattern of 100 $\mu$m diameter disks into the silica. Xenon difluoride was then used to isotropically etch 25 $\mu$m of silicon enabling the undercut silica disk to be reflowed with several watts of focused $\rm{CO}_2$ laser power\cite{Armani03}. The microtoroids  then typically had major and minor diameters of 60~$\mu$m and~6~$\mu$m, respectively, and an undercut of $10~\mu$m between the edge of the toroid and the pedestal. An external cavity diode laser (New Focus TLB-6320) that could be tuned over several microtoroid free spectral ranges provided an in-fiber optical {\it probe field} at 980~nm. The probe field was evanescently coupled into a microtoroid using a single mode tapered optical fiber held within several micrometers of the microtoroid surface with a three axis piezo-driven flexure stage.

We characterize the optical quality factor ,$Q_o$, by fitting the observed transmission spectra to a coupled cavity model\cite{Bowen2003,Gardiner1985}. 
The intrinsic quality factor was found to be $Q _o\! = \! 3 \! \times \! 10^6$ enabling thermal locking of the microtoroid to the probe laser. This thermal locking is a result of the temperature dependent refractive index of silica which with optical heating shifts the microtoroids resonant frequency to remain locked to the probe laser\cite{Carmon}.
All experiments were carried out at room temperature ($T=300$~K) and atmospheric pressure.

To achieve dielectric gradient force actuation a non-uniform electric field was produced with a RF voltage from a network analyzer (Hewlett Packard 3577B) and applied to a sharp stainless steel electrode.  This \emph{sharp electrode} had a 2 $\mu$m tip diameter and was positioned 15 $\mu$m vertically above the center of the microtoroid, with the height chosen to balance the benefit of increased gradient forces against the risk of chemical contamination and structural damage due to physical contact. Stepper motor drives (Thorlabs DRV001) enabled both accurate positioning and fast scanning of the sharp electrode. The silicon wafer with integrated microtoroids was placed between this electrode and a large \emph{flat electrode} that was grounded to a voltage supply but could be offset for particular measurements. A schematic of the electric fields produced by our electrode arrangement can be seen in  Fig.~\ref{expt}B where the method of images, typically used to describe boundary conditions in electrostatics, is used to predict the form of the electric
field lines. Figure~\ref{expt}B shows that a voltage applied across the sharp and flat electrodes induces charge separation and the electric field through the microtoroid is well approximated by that of a point charge located at the tip of the sharp electrode. This was confirmed with finite element modeling. As will be shown later, our experiments suggest that the intrinsic polarization of the silica microtoroid material due to static electric fields dominates the induced polarization due to our RF actuation. This induced polarization is most probably caused by charges trapped in the silica. The inhomogeneous electric field acts on the polarized silica creating the actuation force to stimulate the mechanical modes of the microtoroid.

\section{Results}
\subsection{Ultrasensitive mechanical transduction}
\begin{figure}[b]
\begin{center}
\includegraphics[width=7.5cm]{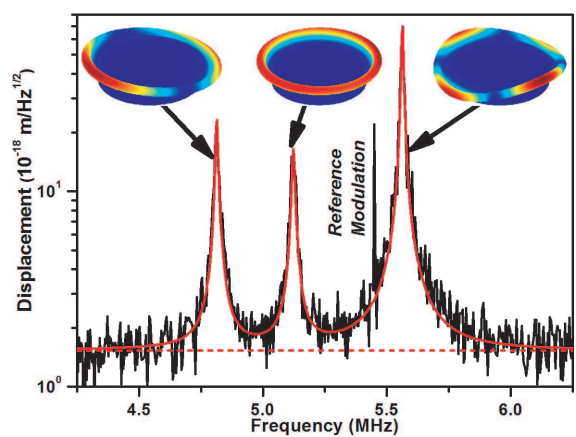}
\caption{(color online) Square-root Brownian motion displacement spectral density. Black curve: measured spectra; solid red (grey) curve: theoretical model; dashed red (grey) line: transduction sensitivity limit. Insets: (from left to right) finite element models of $j=1,2,3$ modes respectively.}
\label{spectra}
\end{center}
\end{figure}

Detuning the probe field to the full-width-half-maximum (FWHM) of the optical mode allowed transduction of the mechanical motion; with mechanical motion induced shifts in the microtoroids optical path length being transferred to the amplitude of the out-coupled optical field and detected directly with a silicon photodiode (New Focus 1801)\cite{Kippenberg2005}. The transduction noise spectral density obtained by spectral analysis of the output photocurrent, without application of an RF drive voltage, is shown in Fig.~\ref{spectra} for 1~mW of incident probe power. By applying a known reference modulation we were able to calibrate the absolute amplitude of the displacement from the optomechanical transduction signal\cite{Schliesser08}. The three spectral peaks in Fig. 2 are  due to the Brownian motion of the microtoroid mechanical modes. The peaks were identified with finite element modeling as the lowest order flexural mode and the two lowest order crown modes, where the frequencies of the modes and the model agree to within 5\%. The flat spectral noise background was due to laser phase noise, as confirmed by an observed square root dependence of its amplitude on probe power. This noise placed an ultimate limit on the transduction sensitivity.

The quantum behavior of a mechanical oscillator can only be observed if the transduction sensitivity is sufficient to resolve the oscillators zero point motion. Brownian motion can be modeled as the mechanical response due to a Langevin force $F_T$ that describes the coupling of the oscillator to the thermal bath with thermal energy $\frac{1}{2}k_B T$\cite{Pinard2000}. The spectral density then has a Lorentzian lineshape\cite{Landau}
\begin{equation}
S_x^{(j)} (\omega)~[{\rm m^2~Hz}^{-1}] \! = \! |\chi_j(\omega)|^2S_{F_T} \!= \! 2 k_B T \Gamma_j m_j|\chi_j(\omega)|^2.
\label{spectraldensity}
\end{equation}
Here, the parameter $j$ identifies each mechanical mode with $j$=1 being the lowest order crown mode, $j$=2 the lowest order radial flexural mode and $j$=3 the second order crown mode; and $m_j$ and $\omega_{m,j}$ are the effective mass and resonance frequency of mode $j$ respectively. The dissipation rate $\Gamma_j$ is the FWHM of the power spectrum and the mechanical susceptibility is given by
\begin{equation}
\chi_j(\omega) \! = \! \frac{1}{[m_{j} (\omega_{m,j}^2 \! - \! \omega^2 \! - \! i \Gamma_j \omega )]},
\label{susceptibility}
\end{equation}
valid when the spectral analysis is at frequencies close to mechanical resonance. In our case the total measured spectral density is the sum of the spectral density of the Brownian motion of each mechanical mode, and a spectral background noise $S_{N}$ due to transduction noise which is approximately flat in frequency
\begin{equation}
S_{\rm total}(\omega) \! = \! \sum_{j=1}^3{S_x^{(j)} (\omega)} \! + \! S_{N}.
\label{totalspectraldensity}
\end{equation}
Using Eqs.(\ref{spectraldensity},\ref{susceptibility},\ref{totalspectraldensity}) it was possible to accurately fit the observed mechanical spectral density as shown in Fig.\ref{spectra}, with $m_j$, $\Gamma_j$, and $S_N$ as free parameters. From this fit we find ${S_N}^{1/2} \! = \!  1.5 \! \times \! 10^{-18}$~m~Hz$^{-1/2}$, and the effective masses and damping rates for each of the three modes shown in Table~\ref{masstable}.
\begin{table}[ht]
\caption{Mechanical mode parameters}
\centering
\begin{ruledtabular}
\begin{tabular}{c c c c c}
Mode & $m_{j}$($\mu$g) & $\Gamma_j/2\pi$(kHz) & $S_{zp}^{(j)}~(\rm{m~Hz}^{-1/2})$ & $S_N/S_{zp}^{(j)}$ \\ [0.5ex]
\hline
j=1 & 280 & 9.5 & $1.4\times10^{-20}$ & 107  \\
j=2 & 410 & 11.5 & $1.1\times10^{-20}$ & 136  \\
j=3 & 33  & 6.8  & $4.6\times10^{-20}$ & 32  \\
\end{tabular}
\end{ruledtabular}
\label{masstable}
\end{table}

The peak of the spectral density of a mechanical mode due to zero-point motion is given by \cite{Schliesser08}
\begin{equation}
S_{zp}^{(j)} \! = \! \frac{\hbar}{m_j \Gamma_j \omega_{m,j}}
\end{equation}
which gives values in the range of $10^{-20}$~m~Hz$^{-1/2}$ for the modes considered here, as shown in Table~\ref{masstable}. Thus our transduction sensitivity is approximately two orders of magnitude away from the zero-point motion as shown in the last column of Table~\ref{masstable}. The best transduction sensitivity achieved is just over thirty times the zero point motion.

Recently a number of techniques have been developed to improve microtoroid based COMS. These include homodyne and polarization spectroscopic techniques that allow the detection of the phase of the light with quantum limited sensitivity, and have recently been applied to  improve the transduction sensitivity for motion transduction in WGM resonators \cite{Schliesser08}. Reduction of the mechanical mode damping rate has been achieved through microfabrication \cite{Anetsberger}. Cryogenic cooling limits noise due to temperature induced fluctuations in the refractive index and placing the microtoroid in vacuum reduces the mechanical damping due to air pressure\cite{Arcizet09}. Mechanical Q factors of 80 000 have been reported with these techniques while retaining ultrahigh optical Q. The COEMS discussed in this paper is compatible with these techniques and their implementation should enable zero-point motion transduction sensitivity.

\begin{figure}
\begin{center}
\includegraphics[width=8cm]{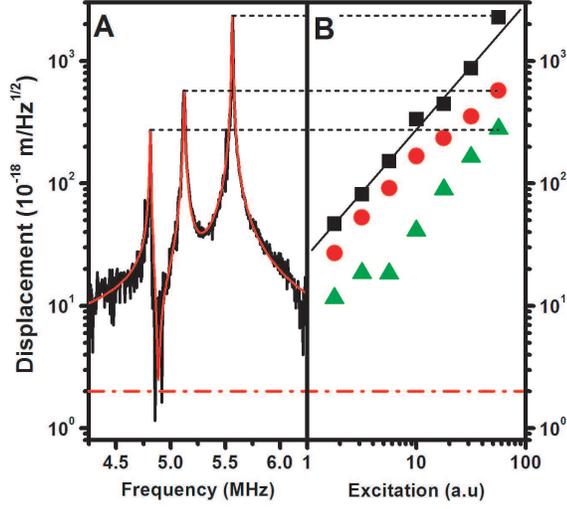}
\caption{(color online) Gradient force actuation of a COEMS. {\bf A} Square root transduction spectral density. Black curve: measured spectra; red curve: theoretical model including interference between the mechanical modes; dash-dotted line: transduction sensitivity limit. {\bf B} Square root peak transduction spectral density as a function of RF drive amplitude. Green triangles: $j=1$, red circles: $j=2$, black squares: $j=3$.}
\label{drive}
\end{center}
\end{figure}

\begin{figure}
\begin{center}
\includegraphics[width=8cm]{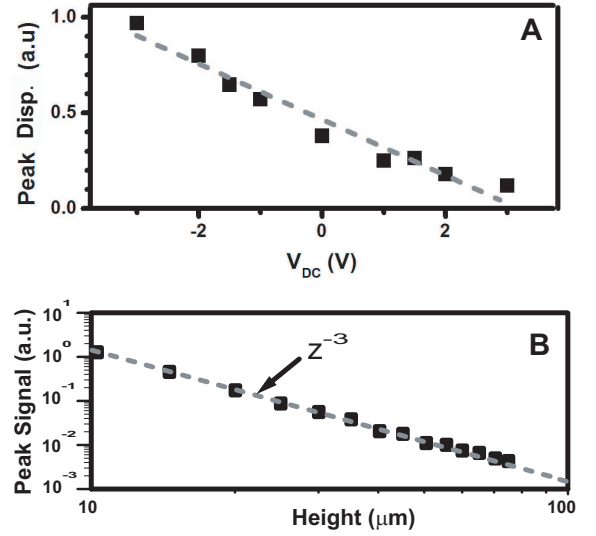}
\caption{(color online) Characterization of the COEMS gradient force. {\bf A} Square root peak transduction spectral density for fixed RF drive amplitude as a function of applied DC voltage. {\bf B} Power spectrum peak as a function of probe height $z$.}
\label{drive2}
\end{center}
\end{figure}

\subsection{Driving mechanical modes}
Electrical gradient force actuation of COEMS offers the possibility of broadband excitation, is compatible with integrated circuits, and is scalable to large architectures. The gradient force actuation of our COEMS was characterized with a network analyzer. The output voltage from the network analyzer was supplied to the sharp electrode and the frequency response was used to monitor the transduction signal. The observed mechanical frequency response is shown in Fig.~\ref{drive}A for a 3 V$_{\rm{rms}}$ network analyzer output voltage. A large increase in the mechanical oscillation was observed as the RF signal was scanned over the mechanical resonance frequency. The inverted part of the lowest order crown mode is due to interference between the two lowest order modes as a result of the coherent driving field. At 3 V$_{\rm{rms}}$ the maximum oscillation amplitude observed was $S^{1/2}_{x,\rm{max}}=2.4\times10^{-14}$~m~Hz$^{-1/2}$ at the peak of the second order crown mode ($j$=3). The peak to peak gradient force can be derived from this measurement in the following way. The dynamics of a damped driven harmonic oscillator can be modeled by the equation of motion
\begin{equation}
m_j\ddot{x}-m_j\Gamma_j\dot{x}+kx=f_o\cos(\omega_jt)
\end{equation}
where $x$ is the displacement of the oscillator from equilibrium, $k=\omega_j^2m_j$ is the spring constant, and $f_o$ is the actuation force. Taking the Fourier transform and rearranging we find
\begin{equation}
f_o=\frac{-i\sqrt{2}m\Gamma_j\omega_jx(\omega_j)}{\sqrt{\pi}}
\end{equation}
The root-mean-square force is then
\begin{equation}
f_{\rm{RMS}}=\langle |f_o|^2\rangle^{1/2}=\frac{2}{\sqrt{\pi}}m_j\Gamma_j\omega_jS_x(\omega_j)^{1/2}\Gamma_{\rm{RBW}}^{1/2}
\end{equation}
where we have used the fact that $\langle x(\omega_j)^2\rangle=S_x(\omega_j)\Gamma_{\rm{RBW}}$. The peak-to-peak force is simply related to the rms force by $F_{pp}=2\sqrt{2}F_{\rm{RMS}}$ so that
\begin{equation}
F_{pp}=\frac{4}{\sqrt{\pi}}m_j\omega_j\Gamma_j\Gamma_{\rm{RBW}}^{1/2}S^{1/2}_{x,\rm{max}}
\label{Fpp}
\end{equation}
For the second order crown mode measured above, we find a peak to peak gradient force of $0.40~\mu$N, surpassing all cryogenically compatible COMS to date by more than an order of magnitude. The linear relationship between the applied RF voltage and the peak oscillation amplitude of each mechanical mode is shown in Fig.~\ref{drive}B where we see the displacement is proportional to the peak drive amplitude (Eq.~\ref{Fpp}).

Gradient force actuation relies on the presence of a net dipole moment. The dipole moment, $\vec{p}$, of the silica in our microtoroid has an intrinsic component due to charges trapped in the silica and an induced component due to the applied DC field, that is 
\begin{equation}
\vec{p}=\vec{p}_{\rm{intrinsic}}+\vec{p}_{\rm{induced}}.
\end{equation}
The electric field of a dipole along the vertical axis $z$ is given by
\begin{equation}
\vec{E}_{dipole}(\vec{r})=\frac{1}{2\pi \epsilon_0}\frac{p_z}{z^3}
\end{equation}
where $p_z$ is the $z$ component of the total dipole moment $\vec{p}$. As the electric field has a $z^{-3}$ dependance, the force exerted by the dipole (the microtoroid) on a point charge (the sharp probe) is expected to also go as $z^{-3}$. Newton's third law tells us that the force exerted by the point charge on the dipole must be the same, and therefore also go as $z^{-3}$. Furthermore, since $\vec{p}_{\rm{induced}}$ is proportional to the applied DC field, the applied force is expected to vary linearly with DC voltage.

The response of our COEMS to applied DC voltage, as predicted above, allowed confirmation that the applied forces were indeed due to gradient force actuation. The peak oscillation amplitude was found to vary linearly with DC voltage (Fig.~\ref{drive2}A) as expected. Furthermore measurements with varying electrode height $z$ confirmed this demonstrating that the oscillation amplitude decays as $z^{-3}$ as shown in Fig.~\ref{drive2}B as predicted. Note that by selecting the DC voltage such that the induced polarization has equal magnitude but opposite sign to the intrinsic component it is possible to zero the $z$-component of the net dipole moment and hence switch off the RF frequency response. By selecting the correct DC voltage can the gradient force can be turned on or off and hence act as a switch in photonic circuits \cite{Tang}.

\begin{figure}
\begin{center}
\includegraphics[width=8cm]{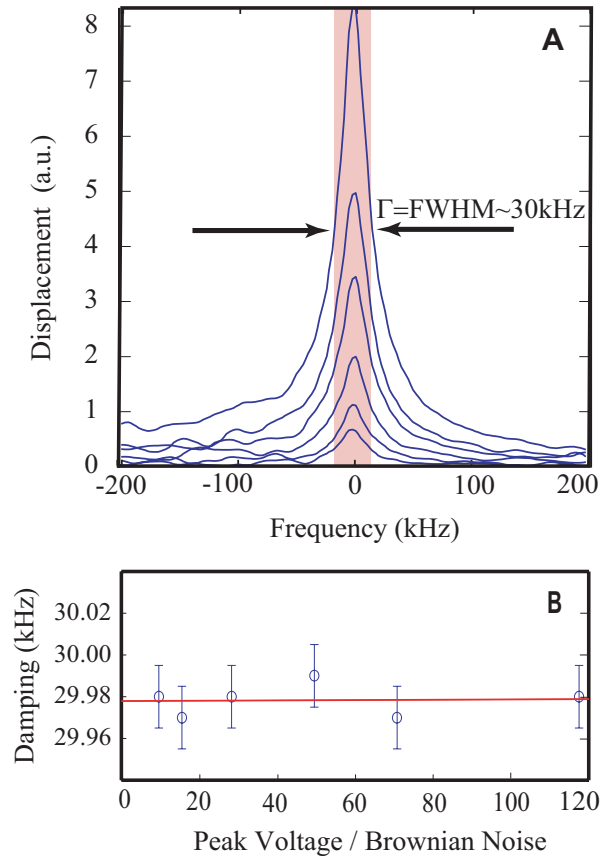}
\caption{(color online) Characterization of the dissipation of the COEMS actuation. {\bf A} Mechanical transduction spectra with varying drive amplitude. The highlighted strip indicates the FWHM. {\bf B} Dissipation rate, $\Gamma$, for each drive amplitude including a line of best fit.}
\label{dissipation}
\end{center}
\end{figure}

There was no observable degradation in the mechanical quality factor of the microtoroid caused by the sharp electrode or the RF drive voltage indicating that the actuation mechanism is low dissipation. To quantify this effect we measure the mechanical dissipation rate at a range of drive voltages for a typical microtoroid mechanical mode. Figure \ref{dissipation}A shows the dissipation rate for peak spectral densities varying from 8 to 120 times the peak of the background Brownian noise. Lorentzian fits allowed the dissipation rate to be extracted in each case, with the results plotted in Fig. \ref{dissipation}B. As can be seen, within experimental uncertainties the dissipation rate is uneffected by the drive voltage amplitude; with the line of best fit to the dissipation rate versus spectral peak amplitude having a slope of 40$\pm$500~mHz/dB

For quantum behavior to be observed in a COMS requires temperatures $T\ll \hbar\omega_{mj}/k_B$. For a mechanical resonance of say 5.5 MHz this requires a very low temperature of 0.26 mK. The ultra-high $Q$ cavities required for radiation pressure actuation experience substantial heating due to intracavity absorption constraining the possibility of reaching the quantum regime. It has been shown that 1 W of intracavity power causes a 10K temperature increase in microtoroids similar to those used in this work\cite{Kippenberg09}. With radiation pressure alone an intracavity power of $P\!\approx\!c F_{\rm rms}/\pi\!=\!36$~W would be required to produce the maximum oscillation amplitude demonstrated in this work\cite{Carmon05}. These power levels therefore clearly cause sufficient heating to be incompatible with the quantum regime. The capacity of our COEMS to achieve transduction sensitivity close to the standard quantum limit and and actuation without appreciable heating should facilitate future quantum communication and quantum nonlinear mechanical experiments.

\subsection{Microscopy of mechanical mode structure}
Displacements in a microtoroid can result from either the propagation of mechanically resonant modes that deform the microresonator structure, as discussed above, or from displacement of the entire resonator due to environmental factors. Movements of the entire structure are not the focus of this paper as they are strongly dependent on how well the microtoroid is isolated from environmental perturbations and usually have low resonance frequencies that can be easily distinguished from the internal resonant modes\cite{Briant2003}. As the quality of the finite element modeling used to predict both frequencies and modeshapes of resonant modes depends heavily on the initial mesh chosen it is important to validate the simulations with experimental results. Here we demonstrate a simple technique to directly measure the driven mechanical modes as previous techniques for measuring mechanical modes\cite{Briant2003,Arcizet2008} are not directly applicable to these whispering gallery mode resonators.

Electrostatic actuation has been previously applied \cite{Arcizet2008} as the actuation force to form a cross sectional image of the mechanical modes of Fabry-P\'{e}rot resonators. Here we apply the electric gradient force to integrated microtoroid cavities, by automatically scanning the sharp electrode across the microtoroid surface while strongly electrically driving the system at  a mechanical resonance frequency and using optomechanical transduction to observe the mechanical response. This allows a two dimensional picture of the resonant mode to be generated.

\begin{figure}
\begin{center}
\includegraphics[width=8cm]{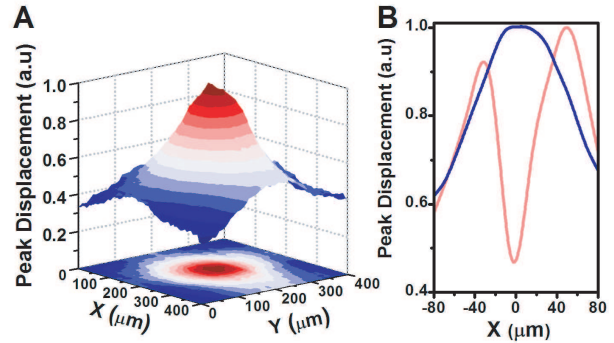}
\caption{(color online) Scanning probe microscopy of mechanical mode structure. {\bf A} Square root peak  transduction spectral density for a radial flexural mode as a function of the lateral position of the sharp electrode. {\bf B} Cross-sections of  the square root  peak  transduction spectral density for a radial flexural mode (dark blue trace) and a second order crown mode (light red trace).}
\label{microscopy}
\end{center}
\end{figure}

The sharp electrode was positioned 15~$\mu$m above the microtoroid surface and
scanned laterally in the XY plane using computer controlled stepper motor drives.
A fixed-frequency drive voltage was applied with frequency chosen to coincide with the peak oscillation frequency of the microtoroid radial flexural mode, and the driven peak oscillation amplitude was observed throughout the scan using opto-mechanical transduction. Consistent with this choice of mode, a radially symmetric profile was found, as shown in Fig.~\ref{microscopy}A.
In contrast, crown modes, whose mechanical motion exhibits sinusoidal oscillations around the circumference of the torus\cite{Schliesser08}, are expected to have a minimum at the microtoroid center, since the forces on out-of-phase nodes cancel; and this was indeed  observed, as shown in Fig.~\ref{microscopy}B. The spatial resolution was limited by the non-point-like nature of the electric gradient force; with the observed structure being a convolution of the actual mechanical mode structure and the spatial distribution of the gradient force. This limitation could be overcome in future by positioning the sharp electrode closer to the microtoroid surface, using electronic control loops similar to those in atomic force microscopy\cite{binning} to avoid unintentional contact.

\section{Conclusions}
The COEMS architecture described in this paper combines the ultrasensitive transduction capabilities of COMS with the gradient force actuation capabilities of NEMS. Forces as large as 0.40~$\mu$N were applied to the mechanical resonator due to the inhomogenous electric field from a sharp electrical probe tip. These actuation forces do not result in significant heating and have significantly higher magnitude than what has been achieved with cryogenically compatible radiation pressure actuation. This COEMS system has a transduction sensitivity of $1.5 \! \times \! 10^{-18}$~m~Hz$^{-1/2}$ on the order of thirty times the mechanical zero-point motion. Lateral scanning of the drive electrode enabled microscopy of the mechanical mode structure, providing a tool to facilitate quantitative understanding of the mechanical properties of optomechanical systems. Measurements of the change in transduction signal with the probe height and applied DC voltage were used to show that the intrinsic polarization of the silica was the origin of the resonators mechanical response to the gradient field. By applying a DC bias voltage it was possible to turn the mechanical resonators response on and off, giving the COEMS a photonic switch capability.

This research represents important progress in our capacity to control cavity optomechanical systems at the quantum level, allowing, for example, feedback cooling and electro-optic spring effects; and presenting an enabling step towards the new regime of quantum nonlinear mechanics, where strong mechanical driving of a ground-state cooled mechanical oscillator allows exploration of quantum nonlinear dynamics and nanofabrication techniques provide the capacity to engineer the nonlinear properties.

This research was funded by the Australian Research Council grant DP0987146, and performed in part at the Australian National Fabrication Facility. We gratefully acknowledge helpful advice  from T. Kippenberg, G. Milburn, and T. Stace; and experimental support from A. Rakic and K. Bertling.

\end{document}